\documentclass[sigconf,authorversion]{acmart} 
 
\AtBeginDocument{%
  \providecommand\BibTeX{{%
    \normalfont B\kern-0.5em{\scshape i\kern-0.25em b}\kern-0.8em\TeX}}}

\usepackage[figuresright]{rotating}
\usepackage{color}
\definecolor{mscolor}{rgb}{0.1,0.6,0.3}
\definecolor{bencolor}{rgb}{0.1,0.1,0.9}
\definecolor{xycolor}{rgb}{0.8,0.2,0.5}
\definecolor{todocolor}{rgb}{1.,0.,0.}

\definecolor{kacolor}{rgb}{0.19, 0.55, 0.91}

\copyrightyear{2024}
\acmYear{2024}
\setcopyright{rightsretained}
\acmConference[CHI '24]{Proceedings of the CHI Conference on Human Factors in Computing Systems}{May 11--16, 2024}{Honolulu, HI, USA}
\acmBooktitle{Proceedings of the CHI Conference on Human Factors in Computing Systems (CHI '24), May 11--16, 2024, Honolulu, HI, USA}
\acmDOI{10.1145/3613904.3642143}
\acmISBN{979-8-4007-0330-0/24/05}

\begin{document}

\title{Design Space of Visual Feedforward And Corrective Feedback in XR-Based Motion Guidance Systems}


\author{Xingyao Yu}
\email{Xingyao.Yu@visus.uni-stuttgart.de}
\orcid{0000-0002-4249-1755}
\affiliation{%
  \institution{University of Stuttgart}
  \streetaddress{Allmandring 19}
  \city{Stuttgart}
  \country{Germany}
  \postcode{70569}
}

\author{Benjamin Lee}
\email{Benjamin.Lee@visus.uni-stuttgart.de}
\orcid{0000-0002-1171-4741}
\affiliation{%
  \institution{University of Stuttgart}
  \streetaddress{Allmandring 19}
  \city{Stuttgart}
  \country{Germany}
  \postcode{70569}
}

\author{Michael Sedlmair}
\email{Michael.Sedlmair@visus.uni-stuttgart.de}
\orcid{0000-0001-7048-9292}
\affiliation{%
  \institution{University of Stuttgart}
  \streetaddress{Allmandring 19}
  \city{Stuttgart}
  \country{Germany}
  \postcode{70569}
}

\renewcommand{\shortauthors}{Yu et al.}

\begin{abstract}
Extended reality (XR) technologies are highly suited in assisting individuals in learning motor skills and movements---referred to as motion guidance. In motion guidance, the ``feedforward’’ provides instructional cues of the motions that are to be performed, whereas the ``feedback’’ provides cues which help correct mistakes and minimize errors. Designing synergistic feedforward and feedback is vital to providing an effective learning experience, but this interplay between the two has not yet been adequately explored. Based on a survey of the literature, we propose design space for both motion feedforward and corrective feedback in XR, and describe the interaction effects between them. We identify common design approaches of XR-based motion guidance found in our literature corpus, and discuss them through the lens of our design dimensions. We then discuss additional contextual factors and considerations that influence this design, together with future research opportunities for motion guidance in XR.

\end{abstract}




\begin{teaserfigure}
  \includegraphics[width=\textwidth]{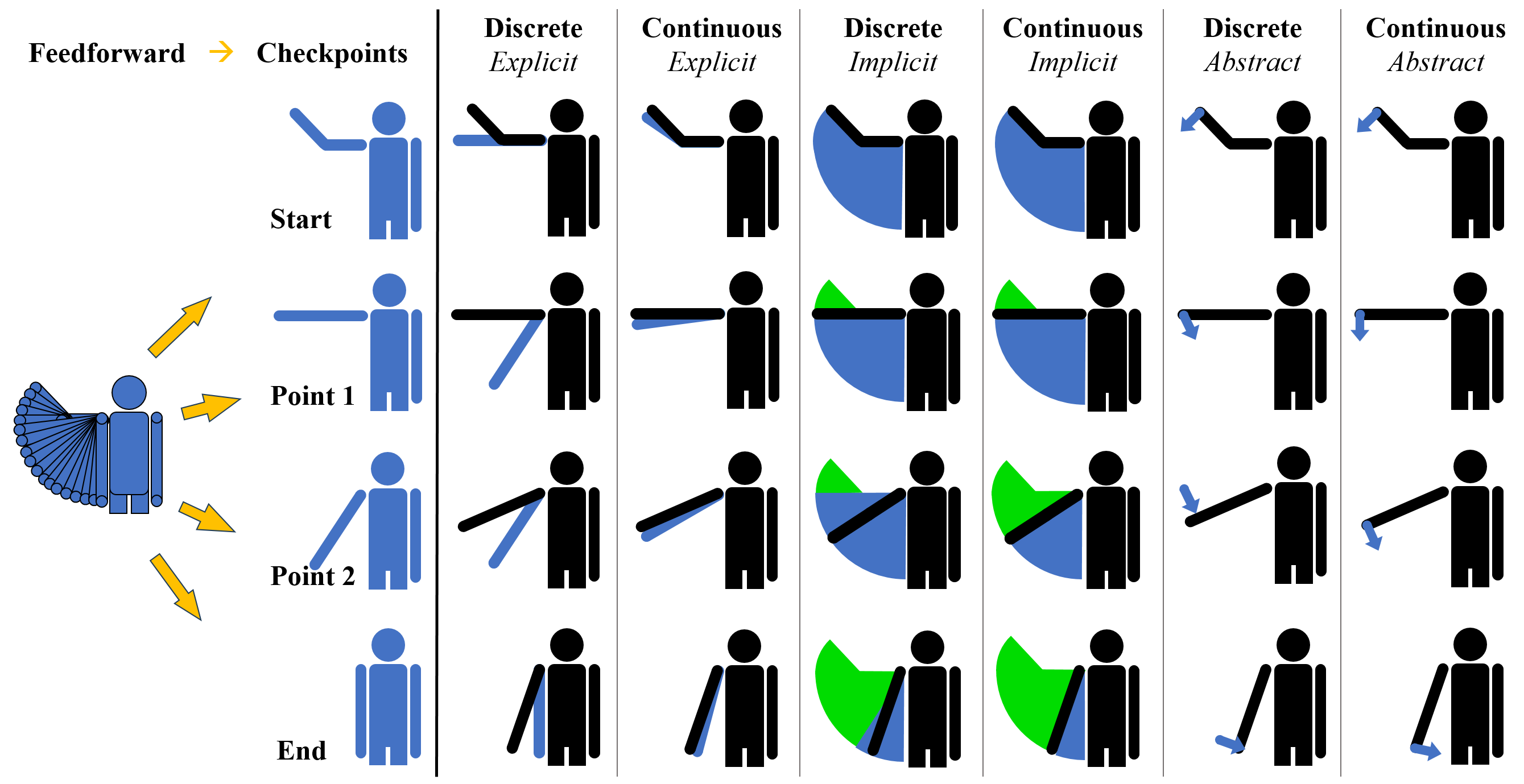}
  \caption{Schematic representations of motion feedforward. The leftmost column shows the complete motion sequence and its critical postures at four critical moments (\textit{Start}, \textit{Point1}, \textit{Point2}, and \textit{End}). The remaining columns show how the feedforward of this motion can be represented using different configurations of \textbf{level of indirection} (\textit{explicit}, \textit{implicit} and \textit{abstract}) and \textbf{interactive update strategy} (\textit{discrete}, \textit{continuous} and \textit{autonomous}). The visualizations in \textcolor{blue}{blue} represent the feedforward; while the the black glyphs show the trainee's four moments during training from top to bottom (\textit{Start}, \textit{reach Point 1, \textit{between Point 1 and Point 2}, and \textit{between Point 2 and End})}; the \textcolor[RGB]{0,220,0}{green} parts visualize the completed parts of a motion when using \textit{Implicit} guidance (column 4 and 5). \textit{Discrete} guidance (column 2, 4 and 6) presents cues for the next \textbf{critical posture}, which updates only when that posture is satisfied. \textit{Continuous} guidance (column 3, 5 and 7) presents cues for the next \textbf{frame} in the motion sequence at all times. \textit{Autonomous} guidance is not shown as it operates independently of the trainee's actions.}
  \label{fig:configuration of feedforward}
  \Description{This is an illustration of the configuration of representation of motion sequence in feedforward. There are 4 rows and 7 columns in this figure. The topmost row provides the name of each configuration while the leftmost column provides an illustration of a complete motion sequence. Each row has its corresponding unique checkpoints, while each column represents a configuration. Thus combining the row and column numbers it is possible to obtain an idea of what the motor feedforward visualization looks like after the trainee has completed the movement at a certain checkpoint. Autonomous guidance is not included in this picture because it is not affected by the trainee's movements.}
\end{teaserfigure}


\maketitle

\section{Introduction}

Augmented Reality (AR) and Virtual Reality (VR), collectively known as Extended Reality (XR), have gained significant attention in recent years for their potential to enhance training and performance in various domains. One particularly promising application of XR is in the domain of motion guidance systems, which assist individuals in learning and achieving specific motor skills, postures, or movements. These systems have the potential to make a positive impact in a wide range of fields, from sports~\cite{chen2022vcoach} and healthcare~\cite{quevedo2017assistance} to industrial applications~\cite{mchenry2020evaluation}. By presenting guidance and feedback around or even on the trainee's body, these systems avoid sloppy or incorrect performance of motions in the absence of instructor supervision~\cite{yu2020perspective}. The complex nature of XR technologies has opened up a vast design space for creating novel and effective guidance systems, as compared to previous technological setups involving flat 2D displays and pointing devices~\cite{bau2008octopocus}. This design space is crucial for addressing key questions that arise when designing XR-based motion guidance systems, such as how the motion guidance is presented, how the system responds to user interaction, and how to help users rectify errors. 

In this paper, we explore the visual design space of XR-based motion guidance systems. A complete guidance system simulates an environment for the trainee to practice or learn physical motions through XR. For this purpose, such a system should encompass two key components: motion feedforward, providing motion instructions before the trainee has taken action; and corrective feedback, aiding in behavior correction and performance improvement in response to the trainee's actions. Hence, we structure our investigation following the same two components. Recently, Diller et al.~\cite{diller2022visual} conducted a related survey to summarize the work of using XR to provide corrective feedback for motor training. However, their survey focuses only on corrective feedback, and does not consider the relationship between it and the presented motion feedforward. Thus, building upon their survey, we consider design dimensions for both motion feedforward and corrective feedback, and explore the potential interplay between them.

As the foundational basis of our work, we curate and review a corpus of 38 papers on XR-based motion guidance systems based on existing literature surveys and following citation trails (Section~\ref{sec:methodology}), forming the foundational basis of our work. For motion feedforward, we discuss design choices in the level of indirection, interactive update strategies, viewing perspectives, and any additional context cues (Section~\ref{sec:design space for feedforward}). For corrective feedback, we discuss design choices in information level, temporality, placement, and presentation methods (Section~\ref{sec:design space for feedback}). Furthermore, we review technical setups of existing XR-based motion guidance systems from the literature to provide insights into the current state of the field with respect to our proposed design space (Section~\ref{sec:existing strategies}). We also illustrate how to utilize our design space to generate the appropriate XR systems based on two different scenarios (Section~\ref{sec:use design space}). We then discuss the broader context, constraints, and factors that influence the design of XR-based motion guidance systems, intending to show how to use our design space under different practical conditions (Section~\ref{sec:constraints}).

The goal of this work is to help create effective and user-friendly XR-based motion guidance systems that can support training, rehabilitation, and skill acquisition across various disciplines. As such, the contributions of this paper are as follows:
\begin{itemize}
    \item Design space for motion feedforward and corrective feedback in the context of XR-based motion guidance.
    \item Analysis of common configurations and designs of motion feedforward and feedback based on a corpus of 38 papers, and identification of multiple underexplored research opportunities in the literature.
    \item Illustrative examples of using the proposed design space to create new XR-based motion guidance systems.
    \item Discussion of the related constraints and factors that further influence the design of motion feedforward and feedback.
\end{itemize}

\section{Background \& Related Work}
\label{sec:related work}

\subsection{XR-Based Motion Guidance Systems}

\label{sec:Definition of MG}
Motion guidance systems provide instructions to assist trainees in performing specific motions in situations where on-site guidance from a human instructor is unavailable. This field covers a wide range of topics, including sports training~\cite{zhang2021watch}, physical rehabilitation~\cite{sousa2016sleevear}, and dance~\cite{gutierrez2022modality}. In this paper, we want to focus on motor skill training and dissociate this work from complex tool handling~\cite{cao2020exploratory}, assembly tasks~\cite{mchenry2020evaluation}, or delicate professional skill training~\cite{martin2023sttar}. Elsayed et al.~\cite{elsayed2022understanding} classified motion guidance into three main groups: (1) \textbf{posture guidance}, which involves providing clear visualizations of target postures at specific frames; (2) \textbf{path guidance}, which focuses on directing trainees' body movements along predefined paths in a 3D space; (3) \textbf{movement guidance}, which offers clear posture visualizations at each progressive frame along the path. In this paper, we will comprehensively explore motion guidance that encompasses all of the aforementioned concepts in the context of XR, and reclassify XR-based motion guidance based on our design dimensions (Section~\ref{subsec:configurations of feedforward}).

To achieve effective training, a motion guidance system needs to adequately inform the trainee what motions to perform and seek to reduce the number of errors made. Thus, it needs to include both \textit{feedforward} (Section~\ref{sec:bg of feedforward}) and \textit{feedback} (Section~\ref{sec:bg of feedback}) mechanisms.    

\subsection{Feedforward}
\label{sec:bg of feedforward}

Feedforward, in a general context, provides users with information about available actions and expected outcomes before an interaction, guiding them on what to do and how to proceed. Muresan et al.~\cite{muresan2023using} describe the three stages of how feedforward operates in a VR environment. First, the user needs to trigger a preview of desired actions. Subsequently, feedforward presents a preview to the user, showing the actions and their outcomes that can be taken at specific locations for specific objects. Then, the user can exit the preview and proceed to perform the action. 

In the field of motion guidance, Sodhi et al.~\cite{sodhi2012lightguide} defined feedforward as components that provide users with information about the shape of a motion before it is executed. This information is oftentimes required for trainees to properly learn and complete specific gesture sequences \cite{bau2008octopocus, freeman2009shadowguides, delamare2016designing}. Feedforward visualization in motion guidance can take many forms, including motion paths~\cite{delamare2016designing}, arrows~\cite{sodhi2012lightguide}, user skeletons~\cite{anderson2013youmove}, hand shadows~\cite{freeman2009shadowguides}, and body outlines~\cite{han2016ar}. Feedforward may also take the form of playful and funny cues to increase engagement, also known as ``serious games'', which are often employed in the field of physiotherapy and rehabilitation~\cite{burke2009serious,da2016mirrarbilitation}. However, in this paper we focus primarily on task-based scenarios that provide clear movement instructions, focusing on the execution of simple, repetitive exercises rather than gamely motivational aspects~\cite{butz2022taxonomy}.

\subsection{Feedback}
\label{sec:bg of feedback}

While feedforward provides guidance on the necessary steps to attain a specific result, \textit{feedback} assists the user in comprehending what has happened or is currently happening. Feedback mechanisms specifically present users with the outcomes of particular interactions that they have made, thus relying significantly on users' past actions and movements within the system~\cite{bau2008octopocus}. 

In the field of motion guidance, there may be two types of feedback: corrective feedback and assessment feedback. Corrective feedback is designed to identify and rectify trainee errors or inaccuracies, thereby facilitating the improvement of knowledge and skills. For motion guidance, corrective feedback is employed when the deviation between the primary object (e.g., trainee's joints~\cite{tang2015physio} or a surgical saw~\cite{martin2023sttar}) and its target position exceeds a ``threshold'' predefined by the trainer or designer~\cite{kallmann2015vr}. 
Corrective feedback can be presented visually~\cite{hoang2016onebody}, tactilely~\cite{jin2013vibrotactile}, auditorily~\cite{chaccour2016computer}, and even multimodally~\cite{schonauer2012multimodal}. While this presents an interesting array of design choices, we focus primarily on visual feedback as it is the baseline mode that almost all motion guidance systems rely on---especially in XR. Regardless of the presentation mode, corrective feedback should be at least capable of indicating when or where errors are occurring or have occurred. In contrast, assessment feedback provides an overall ``score'' to the trainee after each trial is conducted, providing a quantifiable measure of their performance. We discuss this in more detail in Section~\ref{sec:scoring Framework}.

\subsection{Design Space of Motion Guidance}
\label{rw:design space}

Several works have presented design space of how feedforward and feedback can be presented to users. An early work by Bau and Mackay~\cite{bau2008octopocus} proposed a design space for classifying feedforward and feedback mechanisms in gesture training systems. They derived two design dimensions regarding feedforward (level of detail and update rate) and four for feedback (recognition value, filtering, update rate, and representation). Despite using mouse and pen input and not XR, their work is still of high relevance to any motion guidance reliant on vision. Building upon this, Delamare et al.~\cite{delamare2015designing} identified four common design dimensions applicable to both feedforward and feedback mechanisms within gesture guidance systems, encompassing temporal, content, medium, and spatial characteristics. Their work explores the breadth of gesture guidance designs, including gesture recognition and trigger mechanisms for feedforward and feedback, but does not focus on the visual channel---especially in XR. Muresan et al.~\cite{muresan2023using} explored the design of feedforward in VR based on its three key stages (Section~\ref{sec:bg of feedforward}), but focused primarily on physics-based interactions with everyday objects, which involve longer and more complex action sequences compared to motor skill training. 

Most relevant to our work is the survey by Diller et al.~\cite{diller2022visual} of visual corrective feedback cues in XR regarding motion training. They categorized visual corrective feedback in existing motion training designs based on nine attributes: MR technologies, point of view, abstraction type, temporal order, stages of learning, publication venue, body parts, use case, and visual cues. Their classification of visual cues, while suitable for describing common approaches in the literature, does not adequately describe and differentiate between the fundamental properties of each cue. For instance, ``rubber bands'' and ``arrows'' are very similar in that they both indicate a direction, with the main difference being the presence of an arrowhead. In contrast, we believe that the induction of visual cues will contribute to a better understanding of the nature of corrective feedback, and help further explore the relationships between visual cues and other elements within the system. Moreover, their descriptions of corrective feedback also include how future movements are presented (e.g., ``guidance'' in ``abstraction type'' and ``upcoming'' in ``temporal order''), which should instead be classified separately as feedforward. We thereby extend their work to include the exploration of feedforward and corrective feedback in XR separately, the clarification of their functional divisions and connections, and further the analysis of the interplay between these two mechanisms.

In summary, previous work has yet to provide a holistic overview of visual design for XR-based motion guidance. They are either aimed at non-motion guidance scenes, or they do not use XR to present feedforward, and hence lack comprehensive consideration of both feedforward and feedback in the XR context. Our work aims to fill this gap through a literature search combined with our experience in related design. Concretely, we focus on motor skill training with simple and repetitive motions in task-based scenarios, where feedforward and feedback are both presented through XR and mainly on the visual channel.

\section{Methodology of Literature Review}
\label{sec:methodology}

To ensure that our design space is properly grounded in prior XR-based motion guidance research, we decided to conduct a literature review. It is through this process that we could brainstorm, define, and validate our proposed design dimensions for both feedforward and feedback.

To construct our corpus, we employed the snowballing methodology described by Wohlin~\cite{wohlin2014guidelines}. The start set of our snowball comprises three logical subsets. \textit{Start set A} contains two comprehensive survey papers: one on corrective feedback cues by Diller et al.~\cite{diller2022visual}, and one on rehabilitation exercises in XR by Butz et al.~\cite{butz2022taxonomy}. \textit{Start set B} contains prominent publications (over 100 citations on Google Scholar) that fit our scope: Just Follow Me~\cite{yang2002implementation}, ShadowGuides~\cite{freeman2009shadowguides}, LightGuide~\cite{sodhi2012lightguide}, YouMove~\cite{anderson2013youmove} and Physio@Home~\cite{tang2015physio}. \textit{Start set C} contains publications within the past 5 years on XR-based motion guidance across a broad research spectrum, including: Kodama et al.~\cite{kodama2023effects} on training using virtual co-embodiment, Zhou et al.~\cite{zhou2022movement} on motion guidance with an MR mirror, Lilija et al.~\cite{lilija2021correction} on correction on virtual hand avatar movements, Yu et al.~\cite{yu2020perspective} on the influence of perspective in motion guidance, and D{\"u}rr et al.~\cite{durr2020eguide} on the virtual appearance of feedforward.

To expand the corpus, we employed the backward snowballing method on \textit{start set A}, while concurrently applying both forward and backward snowballing methods on \textit{start set B} and \textit{start set C}. For each paper to examine, we first conduct a preliminary screening based on its title, abstract and keywords, and then decide on inclusion or exclusion by reading its system design and comparing it with our intended scope. After including a paper in our corpus, we applied both forward and backward snowballing, and repeated this process until no new papers in its associated citation chain were left. We only included papers published after 2002, which was the year when the Just Follow Me system~\cite{yang2002implementation}, the first system to employ VR for motion training, was published.

We opted against using a database query approach for two main reasons. First, there exist many synonyms for motion guidance, thus making it challenging to search by. For example: motion, motor, movement, and exercise; and guidance, tutorial, training, and assistance. Second, a database query may inadvertently exclude XR-based motion guidance in specific application domains such as medical and sports science, especially when limiting the search to certain publishers and/or venues (e.g. ACM, IEEE).

Our final corpus consists of 38 papers, which includes the papers in our start set. Among them, 16 were published in venues related to XR and visualization (e.g., IEEE VR, ISMAR, TVCG; ACM VRST; Springer VR), 15 in human-computer interaction (e.g., ACM CHI, UIST, IUI, ICMI, ITS, ISS, AH; NordiCHI), and the remaining 7 in other domains such as gaming (IEEE Trans. Games), artificial intelligence (Frontiers in Robotics and AI; ACM TIST), and communication technology (Frontiers in ICT). Table~\ref{tab:existingStratiges_motionGuidance} shows all of the papers in our corpus. While this corpus might not be exhaustive, we believe that it is nonetheless a representative sample of the space.

We use this corpus to devise design space for motion guidance in terms of its two components: motion feedforward (Section~\ref{sec:design space for feedforward}) and corrective feedback (Section~\ref{sec:design space for feedback}). We first extracted initial design dimensions from existing literature~\cite{bau2008octopocus,delamare2015designing,diller2022visual}. We then augmented these dimensions by examining the many different technical setups present in our corpus, drawing on our own experience in the fields of XR and motion guidance. We iteratively refined the design dimensions through many discussions among the authors, aiming to avoid omissions, ambiguity, and overlap between dimensions. Validation was performed at each stage by applying the design dimensions to all papers in our corpus until a final design space was agreed on.

\section{Design Space for Motion Feedforward}
\label{sec:design space for feedforward}
In a motion guidance system, the feedforward provides information on how to move or perform specific motions. It is the instructions, cues, or guidance given to trainees \textit{prior} to the course of a movement, which can help them understand the desired motion and how to execute it properly, such as the motion path~\cite{bau2008octopocus} or the visualization of postures~\cite{freeman2009shadowguides}. We further refine the design space of feedforward in XR-based motion guidance based on the two original dimensions defined by Bau and Mackay~\cite{bau2008octopocus}: level of detail (the granularity of the motion sequence presented in feedforward), and update rate (the frequency that feedforward updates). We adapt these two dimensions to XR as follows. First, the immersive and embodied nature of XR allows motion information to be represented in a very direct and life-like form as compared to on a desktop in Bau and Mackay's~\cite{bau2008octopocus} work, which we capture as the \textbf{level of indirection} of the feedforward. Second, the highly interactive nature of XR has us consider how the feedforward might respond to the trainee's actions throughout the motion, rather than being an independently updating entity, thus resulting in our \textbf{interactive update strategy} dimension. We also consider the broader scope of possibilities to visually present information in XR, which includes both the \textbf{viewing perspective} between the trainee and the feedforward, and any \textbf{additional contextual cues} that are associated with the motion. We further discuss the interactions and possible configurations between these dimensions.

\begin{figure*}
\centering
    \includegraphics[width=\textwidth]{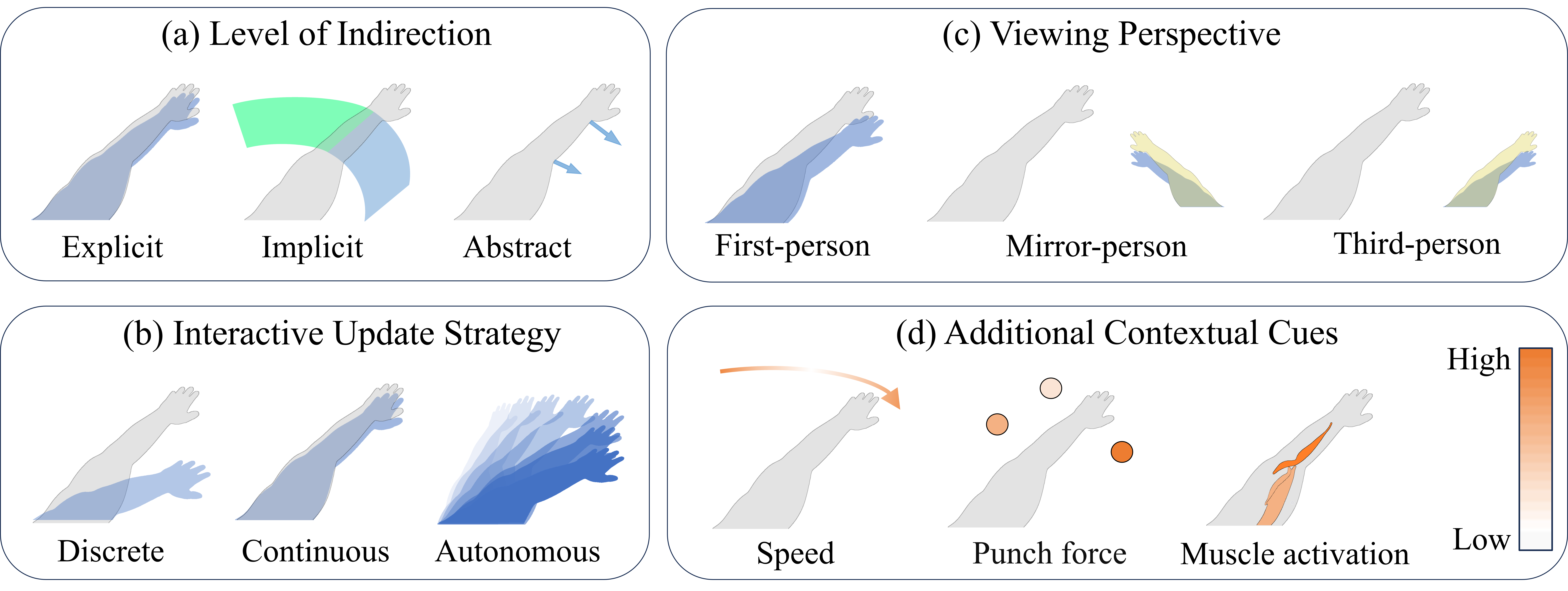}    
    \caption{Design dimensions of motion feedforward. (a) \textbf{level of indirection} at \textit{explicit}, \textit{implicit} and \textit{abstract}. (b) \textbf{interactive update strategy} at \textit{discrete}, \textit{continuous} and \textit{autonomous}. (c) \textbf{viewing perspective} at \textit{first-}, \textit{mirror-} and \textit{third-person} perspectives; in the latter two perspectives, the feedforward will show around the duplicate of the trainee's avatar. (d) \textbf{additional contextual cues} such as \textit{speed}, \textit{punch force} and \textit{muscle activation}. For illustrative purposes, \textcolor[RGB]{90,90,90}{grey} represents the trainee's avatar in an egocentric perspective, \textcolor{blue}{blue} represents feedforward instructions, \textcolor[RGB]{11, 214, 66}{green} represents the completed part in \textit{implicit} guidance, \textcolor[RGB]{227,215,95}{yellow} represents the dynamic duplicate of trainee's avatar in \textit{mirror-} or \textit{third-person} perspective, and \textcolor{orange}{orange} represents severity of speed, punch force, or muscle activation.} 
    
    \Description{This is an illustration of all components in the design space of feedforward. (a) level of indirection at explicit, implicit and abstract. (b) interactive update strategy at continuous, discrete and autonomous. (c) viewing perspective at first-, mirror- and third-person perspectives. (d) additional contextual cues at speed, punch force and muscle activation.}
    \label{fig:all feedforward}
\end{figure*}
\subsection{Level of Indirection}

XR is capable of bringing information into forms that are either inherently abstract, or forms that bear a more direct resemblance to reality~\cite{lee2020data}. The same is true for motion guidance, as the feedforward is capable of either showing exactly what motion should be performed, or instead an (intentionally) vague representation of it. We refer to this as the \textbf{level of indirection} of the feedforward's visual representation. We describe three levels, but note that this should be treated more as a spectrum with no strict boundaries between them (Figure~\ref{fig:all feedforward}a).

\paragraph{Explicit.} The feedforward uses a low level of indirection such that it explicitly represents the desired motion with little to no ambiguity. Usually represented as a 3D graphic, it enables the trainee to simply observe the feedforward and imitate its motion or posture, thus providing no room for misinterpretation. For instance, AR-Arm~\cite{han2016ar} uses transparent 3D arms which provide direct instruction for the trainee to simply overlap their arms with the feedforward. Explicit feedforward need not be realistic in appearance however. For instance, YouMove~\cite{anderson2013youmove} uses stick figures as its feedforward to show full-body postures. It is still obvious to the trainee that they should move and align their own skeletal posture with the feedforward. Therefore, a low cognitive cost is associated with explicit guidance, as trainees can simply follow what they see.

\paragraph{Implicit.} The feedforward uses a medium level of indirection such that it still shows the desired motion, but requires some amount of interpretation and inference by the trainee to accurately perform. A classic example of this is motion guidance systems which use trajectories to show movement paths that need to be followed~\cite{sousa2016sleevear,tang2015physio,yu2020perspective}. While the trajectory, when seen in its entirety, does not present some motion or posture to imitate, it is the traversal from one end of the trajectory to another that results in a guided motion. The need to interpret the feedforward may also be beneficial, as it provides the trainee with more freedom in how they perform the motion. For instance, Kosmalla et al.~\cite{kosmalla2017climbvis} use a projector to highlight hand holds while rock climbing. While this indicates which holds to move their limbs to, it does not indicate \textbf{which} limb to move, thus allowing for more self-decision-making. This decision-making naturally incurs a higher cognitive cost as trainees need to think for themselves the best way to make the desired motion.

\paragraph{Abstract.} The feedforward uses a high level of indirection such that it loosely describes the desired motion, but does little to show how it exactly should be performed. This results in a high level of ambiguity in which the trainee needs to approximate the motion as best they can. An example of this is LightGuide~\cite{sodhi2012lightguide} which projects visual cues onto the trainee's hand such as directional arrows. These arrows, in particular, hint at the trainee a general direction in which their hands should be moved, but with little indication of how far, how fast, or even how straight to move. Alternatively, the trainee may be required to perform the motion purely by memory, with the feedforward serving only as a reminder of the motion to perform. Such use would primarily be to test the trainee's knowledge of already practiced motions, with the lack of guidance being intentional~\cite{lilija2021correction,kodama2023effects}. This clearly has the highest cognitive cost due to trainees not being able to rely completely on the feedforward.

\subsection{Interactive Update Strategy}
\label{sec:interactive update strategy}

For complex 3D motions (e.g., long, intersecting, or overlapping movements), visualizing all postures of the entire motion sequence may be too distracting due to visual clutter~\cite{yu2020perspective}. Hence, motion guidance systems often employ checkpoints throughout the motion sequence, progressively updating or revealing parts of the feedforward in response to the trainee's movement as each checkpoint is reached---when their body aligns with the checkpoint's target posture. Systems can vary in their \textbf{interactive update strategy} (i.e., the distribution of the checkpoints), categorized as either \textit{discrete} guidance, \textit{continuous} guidance, or \textit{autonomous} guidance, as illustrated in Figure~\ref{fig:all feedforward}b.

\paragraph{Discrete Guidance.} Discrete guidance sets checkpoints only for frames related to critical postures. The trainer determines the critical postures in the motion sequence, which trainees should perform as accurately as possible. When the trainee performs critical posture, the checkpoint is met and the next stage of feedforward is shown (and the previous one hidden). This approach can add greater emphasis on crucial moments of the motion, allowing trainees to concentrate on accurately performing the critical postures~\cite{hoang2016onebody}. The effectiveness of discrete guidance is naturally influenced by the choice of critical postures. Alternatively, discrete checkpoints can subdivide a motion into more semantically meaningful components, which may assist trainees in remembering this sequence. For instance, consider a complex ballroom dance sequence with many dance steps performed in rapid sequence~\cite{gutierrez2022modality}. 

\paragraph{Continuous Guidance.} Checkpoints are distributed throughout every single frame of the motion sequence~\cite{han2016ar}, thus providing a constantly updating flow of visual cues that guide the trainee throughout the entire motion. This is a highly responsive form of motion guidance, in which the feedback continuously adapts to the trainee's progress and posture alignment, allowing for real-time adjustments and skill refinement. However, trainees need to visually focus on the constantly updating feedforward, which may lead to a higher dependency on this guidance~\cite{lilija2021correction}.

\paragraph{Autonomous Guidance.} In comparison to the other two strategies, autonomous guidance sees no checkpoints being used, and instead has the feedforward automatically update on its own regardless of the trainee's input. Such guidance is typical of videos or 3D animations~\cite{durr2020eguide,waltemate2016impact}, where trainees can observe the complete motion sequence before attempting it themselves. However, the lack of responsiveness of the feedforward to the trainee's own movements may result in a less engaged learning experience. This may also make it harder to fully grasp the intricacies and nuances of the motion without further context or explanations (e.g., assistive textual cues). However, autonomous guidance is well-suited for presenting motions that need to be performed at a specific speed, be it fast or slow, as the feedforward can show this in real-time.
\\

It is possible to combine multiple interactive update strategies together. For instance, the \textit{GT\_PAUSING} technique by D{\"u}rr et al.~\cite{durr2020eguide} uses both \textit{discrete} and \textit{autonomous} guidance by playing short animations between critical postures of arm motions.

\subsection{Viewing Perspective}

XR research often considers how a user's perspective of the (virtual) world and its content influences their behavior, and this is no different for motion guidance~\cite{yu2020perspective,elsayed2022understanding}. By varying the \textbf{viewing perspective} of the trainee relative to the feedforward, their perception of both the feedforward and of their own body changes, thus influencing their understanding and execution of the desired motions (Figure~\ref{fig:all feedforward}c). We describe three common viewing perspectives in the literature: \textit{first-person}, \textit{mirror}, and \textit{third-person}.

\paragraph{First-Person (1PP)} This perspective tailors the feedforward to be viewed from the viewpoint of the trainee, immersing them directly in the context of the motion, and thereby providing a more natural understanding of it. A common design is to use a transparent ghost arm that the trainee should follow with their own arms~\cite{han2016ar,durr2020eguide}. This encourages a sense of embodiment and kinesthetic empathy, enabling trainees to closely align their actions with the guidance cues. However, motions involving multiple limbs at opposing parts of the trainee's body are not well suited for first-person perspectives, as they would need to frequently turn their heads to see the required feedforwards, thus increasing fatigue~\cite{yu2020perspective}.

\paragraph{Mirror (MPP)} Feedforward can be overlaid on a reflection of the trainee's own body viewed on a (virtual) mirror. Depending on the technological setup, the appearance of such reflections may be consistent with their real bodies (screen-based~\cite{tang2015physio} or MR mirror~\cite{zhou2022movement}), or an avatar~\cite{waltemate2016impact}. The mirror perspective naturally allows the trainee to see their entire body and the feedforward all at once. It can also simulate the trainee's surroundings to provide a meaningful context for a better training experience, such as a wall-size mirror in a dance studio for ballet training~\cite{anderson2013youmove}. 

\paragraph{Third-Person (3PP)} This perspective situates the trainees as an external observer of the feedforward, analogous to watching someone else perform the motion. In XR, this is typically a virtual coach using pre-recorded motion sequences such as to learn Tai-Chi~\cite{han2017MyTai}, but can also be the virtual avatar of a remote trainer providing live motion guidance~\cite{hoang2016onebody}. While distancing the trainee from the motion itself, the third-person perspective provides an exocentric view of the entire motion, thus promoting a more complete understanding of all facets of the motion---especially for motions involving multiple limbs~\cite{elsayed2022understanding}.

\begin{table*}[!t]
    \centering    
    \caption{Possible configurations of motion feedforward and the papers which use them. Illustrations are in Figure~\ref{fig:configuration of feedforward}.}
    \begin{tabular}{c|p{4cm}|p{5cm}|p{4cm}}
    \hline
    Configuration&Explicit&Implicit&Abstract\\\hline
    Discrete&
    \textbf{DisExp}: Explicit representation of desired motions, which updates at critical frames of the motion sequence. Commonly used for posture guidance, with less emphasis on the movements between each critical posture~\cite{caserman2021full,chen2022vcoach,elsayed2022understanding,freeman2009shadowguides,gutierrez2022modality,hoang2016onebody,ikeda2019real,kyan2015approach,lin2021towards,oshita2018self,quevedo2017assistance,turmo2020bodylights,yu2020perspective,barioni2019balletvr,durr2020eguide,anderson2013youmove,schonauer2012multimodal}.&
    \textbf{DisImp}: Implicit representation of desired motions requiring interpretation, which updates at critical frames of the motion sequence. Used for when the trainee can perform the critical posture in any way they wish~\cite{kosmalla2017climbvis}.&
    \textbf{DisAbs}: Approximate representation of desired motions, which updates at critical frames of the motion sequence. Commonly used to indicate the starting point of a motion when testing short-term retention~\cite{kodama2023effects,lilija2021correction}.\\\hline
    Continuous&
    \textbf{ConExp}: Explicit representation of desired motions, which updates throughout the motion when trainees satisfy the postural accuracy. Commonly used for slow, high accuracy motions~\cite{han2016ar,ikeda2018ar,lilija2021correction,kallmann2015vr}.&
    \textbf{ConImp}: Implicit representation of desired motions requiring interpretation, which updates throughout the motion when trainees satisfy the postural accuracy. Commonly used with motion trajectories, where concrete postures are interpreted as the boundaries between completed and uncompleted parts~\cite{yu2020perspective,fennedy2021octopocus,sousa2016sleevear,tang2015physio,gebhardt2023auxiliary}.&
    \textbf{ConAbs}: Approximate representation of desired motions, which updates throughout the motion when trainee satisfies the positional accuracy~\cite{turmo2020bodylights,sodhi2012lightguide}.    
    \\\hline    
    Autonomous&
    \textbf{AutoExp}: Explicit representation of desired motions, which is autonomously played without interaction with the trainees. Commonly used as animations or videos demonstrating the desired motion(s)~\cite{anderson2013youmove,durr2020eguide,escalona2020eva,furukawa2018design,han2017MyTai,hulsmann2019superimposed,kodama2023effects,kosmalla2017climbvis,kyan2015approach,le2019superimposing,mostajeran2019welcoming,sekhavat2018projection,waltemate2016impact,yang2002implementation,zhang2021watch}.&
    \textbf{AutoImp}: Implicit representation of desired motions requiring interpretation, which is autonomously played without interaction with the trainees. A possible example is a motion trajectory that autonomously progresses the boundary between completed and uncompleted parts to indicate the desired speed of motion.&
    \textbf{AutoAbs}: Approximate representation of desired motions, which is autonomously played without interaction with the trainees. A possible example is a sequence of abstract feedforward cues that indicate the order of motions to perform (similar to a \textit{Simon Says} game).\\\hline
    \end{tabular}
    \Description{Possible configurations of motion feedforward and the papers that use them. Illustrations are in Figure 1 and Figure 2.}
    \label{tab:motion feedforward}
\end{table*}

\subsection{Additional Contextual Cues}
\label{sec:velocity vis}

All XR-based motion guidance systems, by definition, should provide feedforward instruction of the motion sequence to perform. In some contexts however, further information relevant to the motion may need to be provided to the trainee. One such information is that of speed. While the feedforward itself, particularly in autonomous guidance, can show the speed outright, \textbf{additional contextual cues} can provide this information instead (Figure~\ref{fig:all feedforward}d). For example, Yu et al.~\cite{yu2020perspective} used a small sphere that moves along the motion path as an indication of speed, whereas Gutierrez et al.~\cite{gutierrez2022modality} encoded the speed of the current ballroom dance segment through the size of solid arrows on the ground. Another form of contextual information is that of muscle activation. This can be seen in the work by Zhu et al.~\cite{zhu2022musclerehab}, which overlays colors on a virtual coach's muscles to indicate to the trainee which muscles are being activated in leg rehabilitation exercises. Of course, it is up to the trainer to decide what additional information they present to the trainee and what cues they use, as this may range from the aforementioned speed and muscle activation, to even the facial expressions to make or the lyrics to be sung at different parts of a dance routine.

\subsection{Configurations \& Commonalities} 
\label{subsec:configurations of feedforward}

Among the four dimensions described in this section, \textbf{level of indirection} defines how to visualize motion cues at a certain moment, while \textbf{interactive update strategy} determines how this visualization will change with the progression of the training. Therefore, these two dimensions collectively present the fundamental information of the motion sequence, giving rise to nine potential configurations of motion feedforward. These configurations are capable of describing all of the feedforward designs in our literature corpus. Descriptions of each configuration and the papers that use them are presented in Table~\ref{tab:motion feedforward}. In addition, Figure~\ref{fig:configuration of feedforward} provides an illustrative view of how \textit{continuous} and \textit{discrete} guidance can respond to trainee interaction (\textit{autonomous} is excluded due to its independent operation from trainee interaction). To further verify the orthogonality of the two dimensions, we developed basic prototypes of each of the 9 configurations to demonstrate how they may function. Videos showcasing these prototypes are included in the supplementary material.

\begin{figure}[b]
    \centering
    \vspace{-4mm}
    \includegraphics[width=\linewidth]{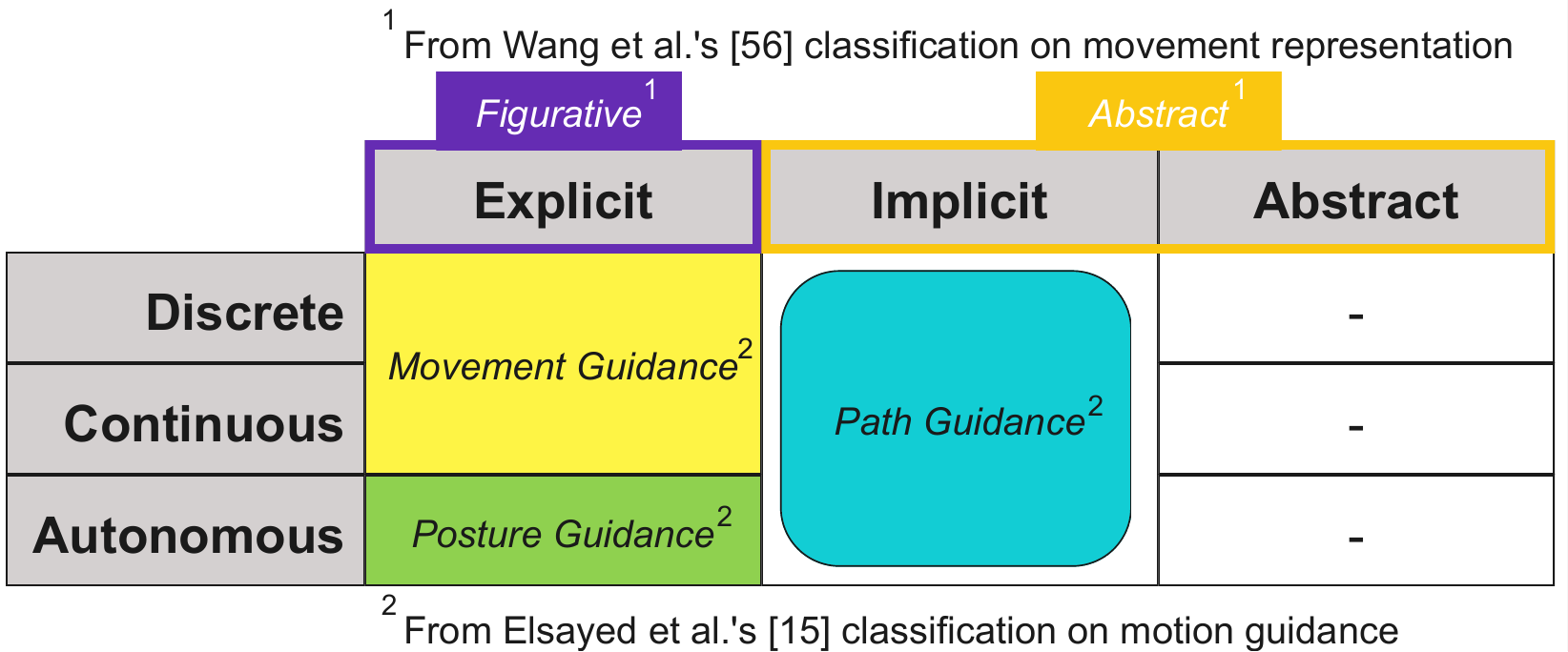}
    \caption{Diagrammatic overview of how our design dimensions of \textbf{level of indirection} and \textbf{interactive update strategy} for feedforward relate to the original classifications by Wang et al.~\cite{wang2022survey} and Elsayed et al.~\cite{elsayed2022understanding}.}
    \label{fig:configurations}
    \Description{This figure explains how our design space of feedforward can cover all existing works. When we compare our work with Wang's work [56], their "abstract" representation corresponds to our implicit and abstract level of indirection and their "figurative" representation should be covered by our explicit feedforward. On the other hand, when we compare our work with Elsayed's work [15], their "posture" guidance refers to our Discrete Explicit, their "path" guidance refers to a subset of our implicit guidance, and their "movement guidance" refers to a combination of Continuous Explicit and Autonomous Explicit.}
\end{figure}

We further validate our nine configurations by comparing them with two existing classifications of motion guidance and representations, and show how they overlap with each other. This is shown graphically in Figure~\ref{fig:configurations}. First, the classification of rehabilitation movement representation by Wang et al.~\cite{wang2022survey}. Their ``abstract'' representation contains motion paths and directional cue representations such as arrows, which correspond to our \textit{implicit} and \textit{abstract} levels of indirection respectively. Their ``figurative'' representation refers to movements presented as a humanoid body, which should be covered by our \textit{explicit} feedforward. Second, the classification by Elsayed et al.~\cite{elsayed2022understanding} on different types of motion guidance, already described in Section~\ref{sec:Definition of MG}. In it, ``posture guidance'' refers to \textit{discrete} \textit{explicit}, ``path guidance'' refers to a subset of \textit{implicit} guidance, and ``movement guidance'' refers to a combination of \textit{continuous} \textit{explicit} and \textit{autonomous} \textit{explicit} guidance.

\section{Design Space for Corrective Feedback}
\label{sec:design space for feedback}

In a motion guidance system, the corrective feedback provides alerts, reminders, and advice to trainees about the errors they make during training. To better investigate this space, we classify corrective feedback into two major groups: \textsc{intuitive} and \textsc{additional}.

\paragraph{\textsc{Intuitive}} \textsc{Intuitive} feedback is borne out of the design of the feedforward. Feedforwards that are \textit{explicit} (and sometimes \textit{implicit}) provide a low enough \textbf{level of indirection} such that trainees can immediately notice disparities between their movements and the feedforward. This binary perception of errors therefore serves as an instinctual indication that the motion is not performed correctly, even in lieu of a quantification of this deviation. Hence, this phenomenon has been regarded as a form of corrective feedback within a number of related literature. In the survey by Diller et al.~\cite{diller2022visual}, the visual cues classified as ``end position'', ``transparent target avatar'', ``opaque target avatar'', ``video overlay'', ``body outline'', ``movement abstraction'' and ``trajectory'' all fall into this group, with the first five cues manifesting themselves through postural differences, while the last two reflecting trajectorial differences. Likewise, systems that use \textit{continuous}~\cite{han2016ar} or \textit{discrete}~\cite{freeman2009shadowguides} guidance naturally indicate errors due to the need to align the trainee's body to the required checkpoint. If the alignment criterion is not met, the feedforward will not visibly progress, thus prompting the trainee to realize that an error has occurred and that they should correct it. Conceptually, these two facets harmonize with each other, forming a symbiotic relationship. 

\paragraph{\textsc{Additional}} \textsc{Additional} feedback can also be employed to overcome some of the limitations that \textsc{intuitive} feedback has. For example, trainees may not even notice that their body is misaligned with the feedforward, particularly in multi-limb motions or errors made in the depth direction~\cite{elsayed2022understanding}. Therefore, considering \textsc{additional} corrective feedback (visual cues like color and arrows) can still be beneficial, which requires additional visual encodings.

We build our design dimensions based on the survey by Diller et al.~\cite{diller2022visual}. First, to remove the ambiguity about feedforward from our design space of feedback, as mentioned in Section~\ref{rw:design space}, we exclude the feedforward elements from their attributes, namely ``upcoming'' from ``temporal order'' and ``guidance'' from ``abstraction type''. Then, we take ``temporal order'' as our \textbf{temporality (when)} dimension, and we include their ``abstraction type'' and ``visual cues'' as our basis to explore the \textbf{presentation (how)} dimension. Finally, we propose dimensions of \textbf{information level (what)} and \textbf{placement (where)} to complete the description of corrective feedback visualization. We believe that this structured division avoids overlap among dimensions, and that assigning descriptiveness to dimensions facilitates a more detailed analysis of feedback and the relationship between feedback and feedforward.

In contrast, since \textsc{intuitive} feedback is presented through the visual differences between the trainee's movement and feedforward rather than visual encoding, its design choices are limited in the \textbf{what}, \textbf{when}, \textbf{where} dimensions. Therefore, we summarize its visual designs only in the \textbf{how} dimension.

\subsection{What: Information Level} 
\label{sec:feedbackLevel}

The first aspect of corrective feedback is \textbf{what} information it tells the trainee, which naturally influences the ease and extent to which they can make corrections. In motion guidance, corrective feedback typically falls under three progressive levels of information: \textit{detection}, \textit{magnitude}, and \textit{rectification}.

\paragraph{Detection.} As the foundational level of corrective feedback, detection provides a straightforward binary output, conveying whether the trainee's body conforms to the required motion or not. Existing works show various approaches in how to encode this output, such as text to display the alignment status of each of the trainee's skeletal joints~\cite{caserman2021full}, or color on the feedforward itself switching between red and green~\cite{hoang2016onebody}. This information level requires the trainer to determine the ideal threshold which dictates the binary switch. Too low a threshold increases the demanded accuracy but is prone to ``flickering'', too high a threshold decreases the demanded accuracy but may make for sloppier trainee movements. Moreover, detection does not provide any further detail about the error or targeted improvements to be made.

\paragraph{Magnitude.} The second level of corrective feedback, quantifies the magnitude and extent of the deviation from the target posture, referred to as deviation distance. This distance is often mapped to visual channels that allow trainees to pre-attentively notice subtle mistakes in their moment, such as transparency~\cite{ikeda2018ar}. Magnitudes can also be seen used in post-hoc analysis after the training process, with the extent of errors being marked on the motion path~\cite{sousa2016sleevear,sekhavat2018projection}. This information level requires the designer to determine the output visual ranges to map the deviation distance to. That is, at what extents does the given deviation distance result in the lowest and highest intensity of visual output.

\paragraph{Rectification.} As the third and highest level of corrective feedback, rectification provides trainees with direct guidance and instructional cues to correct and refine their motions. In other words, it tells the trainee ``how'' to correct their mistakes. For instance, Yu et al.~\cite{yu2020perspective} used lines connecting the trainee's arm to the corresponding target positions on the feedforward. These lines serve a dual purpose: their length indicates the deviation distance, and their direction cues the trainees as to how to adjust their arm to correct the deviation. Alternatively, rectification information can be conveyed by the textual suggestions after the training process. For instance, VCoach~\cite{chen2022vcoach} gives comments to trainees about their performance and suggestions regarding punch direction and speed.

\subsection{When: Temporality} 

Corrective feedback can differ in terms of \textbf{when} it is actually presented to trainees. This includes two main choices: \textit{real-time} and \textit{post-hoc} feedback.

\paragraph{Real-Time.} This is the instantaneous delivery of feedback to trainees as they are engaged in the motion training. This allows trainees to promptly identify any errors in their performance and make timely corrections, thereby creating a rapid learning feedback loop. Examples include showing mismatches in postures~\cite{waltemate2016impact} and the immediate highlight of movement errors~\cite{gebhardt2023auxiliary}.

\paragraph{Post-Hoc.} This delays the delivery of feedback until only after trainees have completed a motion trial. While post-hoc feedback lacks the immediacy inherent to real-time feedback, it provides a longer window for self-reflection, allowing trainees to thoroughly analyze their errors and develop effective strategies for improvement. For instance, post-hoc analysis described in~\cite{escalona2020eva} provides trainees with a review of the movement accuracy over time in the form of a graph, aiming to improve the trainee's movement accuracy at corresponding time points.

\subsection{Where: Placement}

\begin{figure*}[!t]
    \centering
    \includegraphics[width=\textwidth]{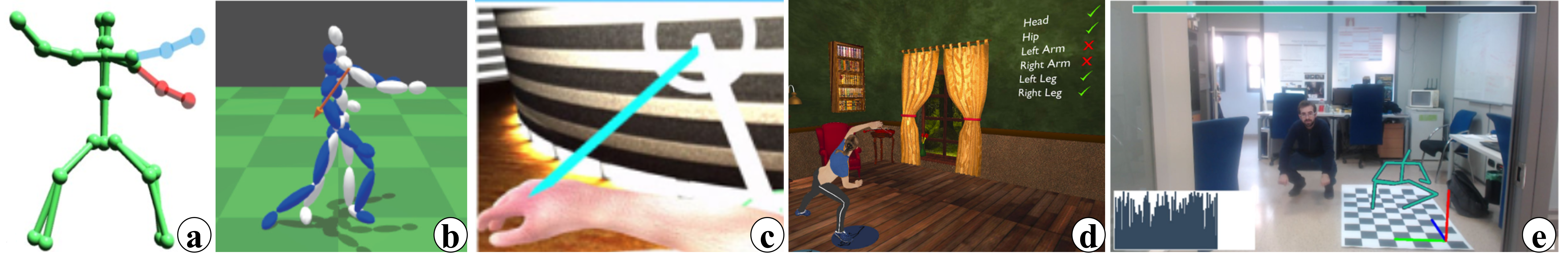}
    \caption{Examples of corrective feedback in the literature:
    (a) \textit{Color}: both the trainee's avatar and the virtual trainer are represented as ball-and-stick models, changing to green when corresponding body parts align, or to red and cyan when they do not align~\cite{hoang2016onebody} (\textit{courtesy of Vetere, \textcopyright ACM}).
    (b) \textit{Direction}: a red arrow is positioned on the \textit{trainee's shoulder} (in white), and provides a \textit{rectification} feedback by pointing at the virtual coach's (blue) shoulder~\cite{oshita2018self} (\textit{courtesy of Oshita, \textcopyright IEEE}).
    (c) \textit{Size}: the length of cyan lines indicates the deviation distance between the trainee's wrist and the desired posture~\cite{yu2020perspective} (\textit{courtesy of Yu, \textcopyright IEEE}).
    (d) \textit{Text}: the \textit{detection} of the alignment status of each body joint attached to a wall~\cite{caserman2021full} (\textit{courtesy of Caserman, \textcopyright IEEE}).\\
    (e) \textit{Graph}: the graph on the bottom-left corner presents the \textit{magnitude} of the movement error throughout the entire training phase~\cite{escalona2020eva} (\textit{courtesy of Escalona, \textcopyright Springer}).} 
    \label{fig:presentation of corrective feedback}
    \Description{This figure shows pictures from existing works to illustrate the visual cues of additional corrective feedback. a) color: the overlapping and non-overlapping parts of the student and virtual instructor's bodies should have different colors. b) direction: a red arrow originated from the trainee's body pointing at its desired position in feedforward. c) size: the line segment connected between the joint of trainee's body and the desired position of feedforward, whose length presents the deviation distance. d) text: the alignment status of each body joint was reported in a text log on the wall. e) graph: the graph shows the movement error magnitude at each frame during training.}
\end{figure*}

\textbf{Where} the corrective feedback is placed directly influences not only the trainee's ability to make corrections, but also their ability to see the feedforward itself. A misplaced or distracting corrective feedback may unintentionally harm and not help the trainee's performance. We outline three strategies of where corrective feedback can be placed that are common in prior research.

\paragraph{Motion Feedforward.} The feedback is embedded directly into the feedforward itself, reducing the need to shift attention between the two. This augments the contextual relevance, enabling trainees to promptly identify errors and refine their motor skills all while visually focusing on the feedforward to execute the motion. An example can be found in the OneBody system~\cite{hoang2016onebody}, where the skeletal joints on the feedforward change color in response to the misalignment of the trainee's body to it (Figure~\ref{fig:presentation of corrective feedback}a). 

\paragraph{Trainee's Body.} The feedback is anchored to the trainee's own body, fostering a heightened connection between the feedback and the trainee's self-embodiment. Here, the ``body'' may be the trainee's own physical body when using AR, virtual avatar while in VR, or the dynamic duplicate in the \textit{mirror-} or \textit{third-person} perspective. For instance, Oshita et al.~\cite{oshita2018self} show how, during post-hoc analysis, arrows embedded into the trainee's avatar can be used to point towards the correct position corresponding to the feedforward. The advantage of this approach is that trainees might perceive the feedback as an extension of their own actions, leading to quicker adjustments in their movements. However, this strategy may introduce visual clutter and distract away from the feedforward~\cite{yu2020perspective}, thus requiring careful design in balancing their saliency while not being overly intrusive (Figure~\ref{fig:presentation of corrective feedback}b).

\paragraph{Environment.} The feedback is projected onto the surrounding environment away from the trainee and the feedforward, thus minimizing undue attentional distractions while performing the motion. Moreover, this frees up the space in the trainee's vicinity for the feedforward to convey a greater amount of information. As an example, Caserman et al.~\cite{caserman2021full} position corrective feedback as text logs on the walls of the virtual environment (Figure~\ref{fig:presentation of corrective feedback}d). However, the challenge lies in ensuring the feedback is still clearly visible, particularly for real-time feedback, as it introduces another visual element that the trainee needs to pay attention to.

\subsection{How: Presentation}
\label{presentation}

Lastly, \textbf{how} the corrective feedback is presented to trainees is important. This refers to the methodological visual representation of movement errors and instructional cues. We draw upon the 13 visual cues summarized in Diller et al.'s survey~\cite{diller2022visual} as a reference, but we also streamlined and consolidated them to explore common visual features and corresponding principles within these cues.

To convey error information and corrective suggestions in a timely manner, corrective feedback should be designed to capture sufficient attention among other visual elements (feedforward, avatar, and surroundings). Therefore, we employ the pre-attentive mechanism~\cite{healey2011attention} (also known as the visual pop-out theory~\cite{munzner2014visualization}) to analyze visual cues. Specifically, we abstract the cues in~\cite{diller2022visual} by the visual features based on their pre-attentive mechanism.

As mentioned at the start of Section~\ref{sec:design space for feedback}, \textsc{intuitive} feedback is borne from the trainee being able to perceive misalignments between the visual feedforward and their own movements at the same moment. Diller et al.'s~\cite{diller2022visual} survey describes several cues that are, we argue, designs of motion feedforward that take advantage of this \textsc{intuitive} feedback as a corrective mechanism. This includes the \textit{shape} of movement for the ``trajectories'' cue, and the \textit{positions} of body joints and the \textit{orientations} of limbs for cues including  ``transparent target avatar'', ``opaque target avatar'', ``video overlay'', and ``body outline''. The differences in these three features can be used to capture the trainee's attention and correct for errors.

The \textsc{additional} feedback mechanism encodes visual attributes of an object to present feedback information. Such an object can be the feedforward representation itself~\cite{waltemate2016impact} (which may serve in conjunction with \textsc{intuitive} feedback), the trainee's avatar~\cite{hoang2016onebody}, or other objects in the environment~\cite{escalona2020eva}. These visual attributes may take the form of \textit{color}, \textit{direction}, or \textit{size}, and may also be reflected in the content of \textit{text} or \textit{graph}. They update as the trainee performs the motion, signifying to the trainee whether the movement is performed correctly.

\paragraph{Color} Color can not only serve as an intuitive indicator of movement accuracy (e.g., green denoting correctness and red denoting errors~\cite{quevedo2017assistance,tang2015physio}), but also enable the representation of the extent of deviations by interpolating color hues. For instance, a higher intensity of red signifies a higher \textit{magnitude} of movement error~\cite{gebhardt2023auxiliary}. 

\paragraph{Direction} Directional cues indicate where and how the trainee's body parts should move to correct for alignment. Both the ``rubber bands'' and ``arrows'' in~\cite{diller2022visual} employ this approach. The rubber band connects the actual and target body positions with a line segment~\cite{yu2020perspective}, while the arrow, when an error occurs, appears at the trainee's deviated body joint pointing towards the corresponding target position~\cite{oshita2018self}. Another example can be found by projecting fan-shaped ripples onto the ground to correct the direction of footstep movement~\cite{sekhavat2018projection}.

\paragraph{Size} Alterations in the length, area, or volume can effectively catch the trainees' attention~\cite{healey2011attention,healey1999large} and evoke a direct association with the \textit{magnitude} of movement errors. For instance, the \textit{length} of the connecting line between body joints of the trainee and feedforward directly visualizes the deviation distance~\cite{yu2020perspective}. Building upon this, Gebhardt et al.~\cite{gebhardt2023auxiliary} replaced this line with a pair of opposing cones, encoding both their heights and base widths in terms of deviation distance between body joints, and eventually presented the movement error by the \textit{volume} of the cones.

\paragraph{Text} Text feedback can serve to alert trainees to the occurrence of errors, such as the mismatch of body joints~\cite{caserman2021full}, or go further by providing direct corrective suggestions~\cite{oshita2018self}.

\paragraph{Graph} 
Graphs provide an abstract method of presenting movement information to the trainees and trainer (e.g., speed and deviation distance). It is well suited, for example, to provide a comprehensive visual overview of movement errors throughout a continuous motion sequence~\cite{escalona2020eva}.

\section{Common design approaches of motion guidance}
\label{sec:existing strategies}

With a corpus of 38 papers on XR-based motion guidance and a design space to describe them, we now delve into the exploration of its common design approaches. We were particularly interested in examining popular design choices and the commonalities between them. As such, Table~\ref{tab:existingStratiges_motionGuidance} shows all 56 motion guidance setups found in our corpus, along with the distribution of their design choices and use cases in our design space. Note that as the design of \textsc{intuitive} feedback is strictly based on the feedforward, we only list design choices related to \textsc{additional} feedback. To achieve a more comprehensive understanding of the literature, we categorized the use cases of all setups into four categories: 
\begin{itemize}
    \item \textcolor{green}{$\blacksquare$} \textit{Sports training}, in which the training process may involve high-intensity movements that may cause fatigue.
    \item \textcolor{yellow}{$\blacksquare$} \textit{Rehabilitation}, which aims at assisting in the recovery and healing of the body.
    \item \textcolor{cyan}{$\blacksquare$} \textit{Dance tutorials}, which aims to provide instruction on various dance techniques. 
    \item \textcolor[RGB]{177,160,199}{$\blacksquare$} \textit{Others}, which contains training that have no requirements for movement speed or have much higher requirements for movement accuracy than temporal performance.
\end{itemize}

After making numerous observations of how feedforward and feedback have been used together in XR-based motion guidance, we also identify potential open research questions and opportunities where relevant.

\paragraph{Autonomous feedforward is typically explicit.} All 19 setups that employed the \textit{autonomous} update strategy exclusively utilized \textit{explicit} guidance. One possible reason is that autonomously played motion feedforward does not wait for the trainee to interact, leading to a shorter duration for the trainee to observe and comprehend individual frames of motion. Therefore, there is a preference for using \textit{explicit} guidance to ensure efficiency in observation and imitation. Based on our description in Table~\ref{tab:motion feedforward}, theoretically, \textit{implicit} and \textit{abstract} guidance can undergo \textit{autonomous} updates, which is technically feasible yet remains unexplored in the literature.

\begin{table*}[htp]
    \centering    
    \caption{56 motion guidance setups assigned to each of our design dimensions from a corpus of 38 papers.}
    \vspace{-0.25cm}
    \hspace{1cm}
    \includegraphics[height=0.97\textheight]{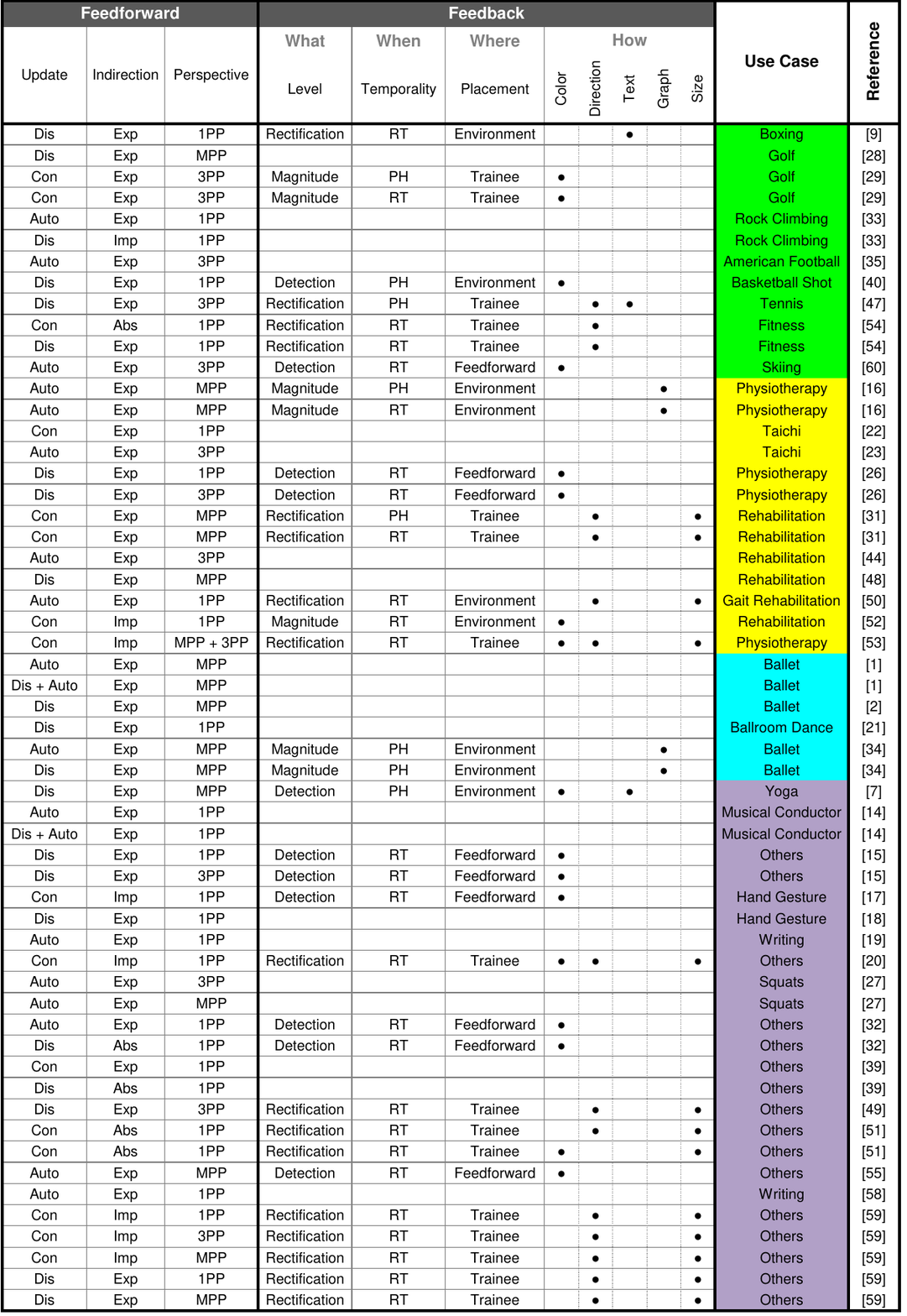}
    \Description{56 motion guidance setups assigned to each of our design dimensions from a corpus of 38 papers.}
    \label{tab:existingStratiges_motionGuidance}
\end{table*}

\paragraph{Use case affects perspective choice.} The frequency of \textbf{viewing perspective} varies across different categories of use cases. In \textit{Sport Training}, \textit{1PP} is the most prevalent, followed by \textit{3PP}, with only one instance of \textit{MPP}. Because these types of movements involve muscle strain or speed requirements, it is necessary to minimize coordinate system transformations related to motion direction---especially symmetric transformations---in order to reduce fatigue. While for \textit{rehabilitation}, the use of the \textit{1PP} is the least frequent (4 out of 13). Compared to \textit{MPP} and \textit{3PP}, \textit{1PP} is generally more intuitive and beneficial for maintaining postural accuracy of 3D complex motions~\cite{yu2020perspective}. However, \textit{rehabilitation} exercises are relatively simple and typically consist of 2D segments, which may not fully leverage the advantages of the \textit{1PP}. Moreover, \textit{MPP} and \textit{3PP} are more commonly employed in traditional rehabilitation scenarios. Among the six cases of \textit{dance tutorial}, five of them employed the \textit{MPP} to instruct ballet in a simulated rehearsal studio, while the only ballroom dance tutorial projected cues of footwork on the ground in \textit{1PP}. Lastly, within the 25 use cases in \textit{others} category, \textit{1PP} is the most frequently used, with a discontinuous style (16 out of 25), as these cases specifically emphasize the postural accuracy of 3D motions. Building upon these findings, we believe that despite the introduction of novel display and interaction techniques, the design of XR-based motion guidance should still consider as much of how trainees learn and practice motions in the traditional real-world context, especially in terms of perspective.

\paragraph{Intuitive feedback is induced by feedforward.} Twenty-one setups do not use any \textsc{additional} corrective feedback. Of these, 19 used an \textit{explicit} form of feedforward, suggesting that a low \textbf{level of indirection} is indeed sufficient to not require further forms of feedback. The two exceptions to this used \textit{implicit} feedforwards. The work by Lilija et al.~\cite{lilija2021correction} focused on testing short-term retention of hand movements, and thus would not require corrective feedback in this context. The other on ClimbVis by Kosmalla et al.~\cite{kosmalla2017climbvis} is primarily motivated by presenting route of holds to follow. In addition, two special cases display corrective feedback associated with the feedforward in ways not described in Section~\ref{presentation}. One case used a progress bar in the \textit{environment} to indicate time spent maintaining a posture, which would reset when this posture is broken, thus indicating a \textit{detection}-level corrective feedback~\cite{barioni2019balletvr,anderson2013youmove}. The second case occurs in LightGuide~\cite{sodhi2012lightguide}, where the \textit{abstract} feedforward and \textit{rectification} feedback are merged together onto an arrow projected onto the trainee's hand, which always points towards the endpoint of the motion regardless of hand position or errors made. 

\paragraph{Non-rectification feedback requires explicit feedforward.} Among the 18 cases where \textit{magnitude} or \textit{detection} feedback was employed, 15 employed \textit{explicit} guidance, two used \textit{implicit} guidance, and one provided \textit{abstract} cue for testing short-term retention~\cite{kodama2023effects}. This was because the \textsc{additional} feedback in these cases did not explicitly instruct how to rectify movements, necessitating reliance on \textsc{intuitive} feedback---the feedforward to correct the trainee's movements.

\paragraph{2D corrective feedback} A small number of setups utilized 2D forms of corrective feedback. \textit{Text} was employed in three setups: in one case it was used as a form of \textit{detection} to indicate unaligned joints~\cite{caserman2021full}, and in the two other cases it was used as \textit{rectification} to provide suggestions for improvement~\cite{chen2022vcoach,oshita2018self}. \textit{Graph} was instead used in four setups found in two papers. These depict feedforward motion sequences over time along the horizontal axis, with data points showing trainee errors at corresponding moments of posture completions~\cite{escalona2020eva,kyan2015approach}. As such, this form of feedback typically provides the \textit{magnitude} or the error. Both \textit{text} and \textit{graph} feedback are all found to be presented as floating panels in the \textit{environment}, and none have attempted to explore alternate ways to place this feedback, such as on the \textit{trainee's body}, although situated visualization has revealed this possibility~\cite{lee2023design}. 

\paragraph{Detection feedback} In the 35 setups that incorporated \textsc{additional} corrective feedback, 11 of them employed feedback at \textit{detection}-level, and they all presented it in the form of \textit{color}. Among these 11 cases, two~\cite{caserman2021full,lin2021towards} presented feedback \textit{post-hoc} in the \textit{environment}, while the remaining nine changed the \textit{color} of \textit{feedforward} in \textit{real-time}. This preference towards the latter is likely because \textit{detection}-level feedback provides only binary information, which is usually presented by an abrupt change in visual cue. In contrast to \textit{direction} and \textit{size}, the abrupt change in \textit{color} does not obscure the \textit{feedforward} and would therefore feel more natural, as they are commonly encountered in daily life (e.g., traffic lights).

\paragraph{Magnitude feedback} Seven setups provided \textit{magnitude} feedback, and 4 of them used \textit{graph}, while 3 utilized \textit{color} hue to represent the deviation distance. 

\paragraph{Rectification feedback} 17 setups featured \textit{rectification}-level feedback. Two papers~\cite{chen2022vcoach,oshita2018self} presented this feedback using \textit{text}, whereas the one by Oshita et al.~\cite{oshita2018self} used both \textit{text} and \textit{direction} to indicate where the limb should be moved to. Apart from this, 14 other setups used \textit{direction}. This preference is understandable as directional cues can clearly guide the correction of body movements. A notable outlier is LightGuide~\cite{sodhi2012lightguide}, which used variations in the \textit{area} of regions with different \textit{colors} to instruct where to move the hand. Moreover, \textit{text} can also supplement this cue by providing suggestions on the strength and speed of movement as well~\cite{chen2022vcoach}.

\paragraph{Size feedback} The occurrence of \textit{size} is highly dependent on \textit{direction}, which is the reason why feedback incorporating \textit{size} is categorized under \textit{rectification} rather than \textit{magnitude}. Among the 13 setups that employed \textit{size}, 12 represented it as the \textit{length} of a \textit{directional} cue; while the last one, LightGuide~\cite{sodhi2012lightguide}, used variations in the \textit{area} of different colors to indicate the \textit{rectification}. 

\paragraph{Potential opportunities for visual features} In Section~\ref{presentation}, we described for \textsc{intuitive} feedback three features (\textit{position}, \textit{orientation} and \textit{shape}); and for the \textbf{encoding} of \textsc{additional} feedback three features (\textit{color}, \textit{direction} and \textit{size}) and two cues (\textit{text} and \textit{graph}). While we based our design space on both the literature and our own experience, there exist unused visual pre-attentive features, and novel uses of previously used features. For example, when utilizing a 3D arrow pointing at the target position to correct a movement error, we can provide additional \textit{magnitude} feedback through its \textit{rotation speed} (unused feature) and \textit{shape} (used only for \textsc{intuitive} feedback), i.e., the greater the \textit{magnitude} of the error, the \textit{faster} it rotates and the greater the \textit{curvature} of the arrow. 

\section{Using The Design Space}
\label{sec:use design space}
We now briefly illustrate how our design space can be used to generate and describe new XR-based motion guidance systems based on two hypothetical scenarios: learning sign language and practicing deadlifts.

\subsection{Scenario 1: Sign Language}
\label{sec:use design space_sign language}
Suppose the British Sign Language (BSL) alphabet is to be taught using XR. As the BSL alphabet comprises a unique hand pose for each letter, the feedforward should be \textit{explicit} to allow the trainee to simply recognize and imitate it. A \textit{discrete} update strategy can help ensure trainees are given enough time to accurately look at and perform each hand pose \cite{hoang2016onebody}. In terms of perspective, the feedforward may either be \textit{MPP} to imitate traditional sign language training with an instructor, or be \textit{1PP} as BSL movements occur entirely in the frontal field-of-view of the trainee, or even a combination of both \textit{MPP} and \textit{1PP} for redundancy.

In this approach, \textsc{intuitive} feedback is already built into the \textit{explicit} feedforward. However, \textsc{additional} feedback may assist the trainee further, especially for complete beginners. This may include the use of color to provide \textit{detection} feedback of misaligned hand and finger joints, or possibly \textit{directional} arrows to guide trainees to \textit{rectify} said misalignments. Whilst \textit{text} could be employed, providing feedback for each individual joint similar to Figure~\ref{fig:presentation of corrective feedback}d may be excessive. An alternative would be to instead provide feedback for the whole posture, such as a binary \textit{detection} that states whether the pose has been achieved or not, or a text description explaining how to \textit{rectify} the mistake \cite{chen2022vcoach}. Lastly, \textit{graph} would not be useful in this scenario as it mainly provides time-varying scalar data, which is not relevant for \textit{discrete} postural guidance.

\subsection{Scenario 2: Deadlifting}
Now suppose XR is used to learn and perform deadlifts, a strength training exercise wherein the trainee is required to lift a barbell from the ground starting from a bent position until they are standing upright, all while maintaining a straight back. This motion involves changes in angles at multiple joints of the body and is vertically symmetrical. Thus, in addition to using \textit{MPP} similar to a physical mirror found in a gym, \textit{3PP} can also present the side-view of the trainee to keep track of body angles in real time---especially of their back. Because deadlifts are a physically intensive exercise, \textit{explicit} feedforward can be used to avoid needless cognitive load. For the same reason, an \textit{autonomous} update strategy can be used, as both \textit{continuous} and \textit{discrete} strategies impose posture checkpoints which may unnecessarily shift the trainee's focus towards satisfying said checkpoints, rather than safely performing the full range of motion in one smooth movement.

This type of training is reliant on the coordination of full-body movements throughout the entire motion. Therefore, in addition to the use of \textit{color} and \textit{arrows} as in Scenario 1, it is also possible to present changes in the key angles of the body which the trainee can study \textit{post-hoc}. This allows for the safe analysis of mistakes whilst not under the physical stress of the exercise. A \textit{graph} can be used to present this information, possibly with an accompanying \textit{text} which provides more actionable suggestions that a personal trainer might typically give (e.g., ``Your waist extended too fast!'').

\section{Discussion}
\label{sec:constraints}

During our review of the literature, we observed other factors that tend to be considered in addition to the aforementioned feedforward and feedback design. These factors are mainly contextual considerations and/or constraints that influence the design of the XR-based motion guidance system as a whole. In this section, we discuss these factors and provide guidelines and implications for future designers and researchers in XR-based motion guidance.

\subsection{Motion Features}
\label{sec:motion features}

A motion sequence can be defined by several features that comprise the movement and determine how it is to be performed. We briefly describe a non-exhaustive list of the main motion features we found in the literature.

\paragraph{Speed.} Certain motion types and training purposes may consider speed as important. For instance, ballroom dancing requires mastery of both footwork and timing~\cite{gutierrez2022modality}, whereas most rehabilitation exercises might only require postural accuracy~\cite{tang2015physio}. Motions involving speed would need to indicate this temporal aspect either \textit{autonomously} or through some \textit{additional contextual cue}. However, we note that the design of the motion feedforward would likely influence the trainee's ability to meet the speed requirements. For instance, Elsayed et al.~\cite{elsayed2022understanding} found that first-person perspectives had a higher angular error in fast movements, but third-person perspectives instead had a lower angular error for these fast movements.

\paragraph{Involved Limbs.} As more limbs are involved in the motion, the demands for body coordination increase, thus making it harder to execute~\cite{elsayed2022understanding}. Not only does the trainee need to pay attention more to their limbs, but the overall spatial area around the trainee which the motion is performed in naturally increases. This, in turn, reduces the overall space that can be dedicated for motion feedforward and feedback, as most of it would need be allocated to the trainee performing the motion itself.

\paragraph{Motion Area.} This refers to the spatial extent and areas in which the motion is executed. This is directly related to the manner which the feedforward is presented: the feedforward should ideally be presented in such a manner that it encompasses the entirety of the motion area within the trainee's field-of-view. As a practical example, consider motions that require both arms to be along the left and right periphery of the trainee's field-of-view~\cite{yu2020perspective}. While the \textit{first-person perspective} requires head turns caused in looking at the feedforward for either arm, the \textit{third-person perspective} allows both to be viewed simultaneously. Such \textit{third-person perspectives} have been found to perform significantly better for motions performed \textit{behind} the trainee as well~\cite{elsayed2022understanding}.

\paragraph{Trajectories.} Many physical motions can be considered a form of spatiotemporal trajectory. As such, many motion guidance systems rely on showing the paths that body joints should follow in space~\cite{sousa2016sleevear,tang2015physio}. Yu et al.~\cite{yu2020perspective} summarized different features of motion trajectories that could influence the complexity of motions: trajectory length, number of intersections, number of arm joints, and the number of 2D segments. These features may induce physical fatigue, limb coordination challenges, and cognitive load for the trainees. While they did not conduct further analysis on these features, highly complex motions would likely require more carefully designed feedforward and feedback representations, taking into account challenges such as visual clutter and overlapping or crossing trajectories.

\subsection{Objective Scoring Framework} 
\label{sec:scoring Framework}

The objective scoring framework is used to provide overall assessment feedback to trainees with respect to postural~\cite{elsayed2022understanding,tang2015physio,lilija2021correction}, temporal~\cite{sodhi2012lightguide,yang2002implementation,lilija2021correction}, or other task-based performance such as basketball shooting accuracy~\cite{lin2021towards} and punching speed~\cite{chen2022vcoach}. It is determined by the training purpose and constrains decisions regarding feedforward visualization. For instance, when there is a requirement for trainees' \textit{temporal} performance, it will be necessary for the feedforward to provide velocity visualization, demonstrating the temporal aspects of the motion to the trainee~\cite{yang2002implementation}. In the case of \textit{postural} aspect, if correctness is calculated only at the key postures, then the \textbf{level of indirection} of the motion feedforward should contain a \textit{discrete} component, to emphasize the significance of specific postures. This is due to the fact that trainees may lose patience and focus during a long motion sequence. However, the constraint imposed by the objective scoring framework on feedforward design is unidirectional. For instance, when feedforward is updated in a \textit{discrete} manner, both key-posture-based~\cite{chen2022vcoach} and frame-by-frame calculations~\cite{yu2020perspective,elsayed2022understanding} can be found in the literature.

\subsection{Individuality} 

While we refer to a ``trainee'' as a hypothetical user, in reality, everyone has unique physiological traits, cognitive abilities, learning styles, and emotional factors that directly impact their progress in motor learning. For instance, Li et al.~\cite{li2018personalized} found that individuality affects image perception. Although not related to motion guidance systems, their work does suggest that varying levels of visual perception between people could influence the effectiveness of the motion guidance system---in this case, one that is reliant on vision. Moreover, the motion guidance system could be tailored to support the trainee's goals, such as by adjusting the difficulty of motions presented to them, or to support their current familiarity with the presented motion, such as by varying the thresholds that determine the \textbf{information level} of corrective feedback, or by increasing the \textbf{level of indirection} to increase the cognitive need for recall of the tasked motion~\cite{lilija2021correction}.

\subsection{Similarity in Appearance} 
As previously alluded to, the choice of technology (i.e., AR vs VR) influences the appearance of the trainee. In the case of AR, it is natural to default to either a human-like avatar, or to just rely on the trainee's own body. In the case of VR, avatars typically take on a wide array of forms, from the same human-like avatars~\cite{durr2020eguide}, to more abstract looking ball-and-stick models of skeletal limbs and joints~\cite{hoang2016onebody}, to bodies formed out of capsules~\cite{oshita2018self}. Previous work by D{\"u}rr et al.~\cite{durr2020eguide} has indicated that using humanoid avatars for the trainee can evoke higher performance regarding movement accuracy as compared to more abstract avatars. There appears however to be a limited body of research addressing a broader spectrum of similarity in appearance between explicit posture guidance and avatars. Given that motion guidance tasks inherently encompass elements related to visual search and matching, there is a plausible basis for asserting that the similarity in appearance between avatars and feedforward cues can influence trainees' performance.

\section{Conclusion}

In this paper, we investigated the visual design of motion feedforward and corrective feedback in XR-based motion guidance system. We first surveyed the literature to understand how researchers have designed such systems and formed our corpus. We proposed four dimensions each for motion feedforward and corrective feedback, then discussed the interplay between feedforward and feedback based on our corpus. We also showed how to use our design space to create new systems using two hypothetical scenarios as examples. We then discussed the additional considerations that stem from the context of the motion guidance.

While outside of the scope of our literature survey, we note that several motion guidance systems consider additional forms of data that could further enhance or improve training effectiveness and outcomes. For instance, biometric sensor information is widely used in rehabilitation guidance (e.g., electromyographic~\cite{melero2019upbeat} or electrical impedance tomography~\cite{zhu2022musclerehab}). As XR technologies develop, it is likely that such data becomes readily accessible to use in any motion guidance system. For instance, eye-tracking may help improve the visibility of both feedforward and feedback cues during complex motions with large \textit{motion areas}, whereas heart rate may support more \textit{individuality} between trainees with different fitness levels.

Finally, as discussed in Section~\ref{sec:constraints}, there are numerous factors that influence the effectiveness of motion guidance systems. While we described how each factor may alter the design choices used with respect to our design space, their effect size and core mechanisms remain unexplored. Thus, the relative importance of each one remains unclear---especially when the context is considered. It may be that the \textit{motion features} governs all other design choices in some contexts, whilst the \textit{objective scoring framework} and the intended learning goal could be prioritized heavily in others. Moving forward, future work could examine how these priorities influence the design of motion guidance systems.

\begin{acks}
This work is funded by Deutsche Forschungsgemeinschaft (DFG, German Research Foundation) under Germany's Excellence Strategy - EXC 2075 – 390740016. We acknowledge the support by the Stuttgart Center for Simulation Science (SimTech).
\end{acks}
\balance
\bibliographystyle{ACM-Reference-Format}
\bibliography{bibliography}










\end{document}